\documentclass[10pt, conference, letterpaper]{IEEEtran}
\pdfoutput=1
\usepackage{diagbox}
\usepackage{graphicx,epsfig,cite}
\usepackage{amssymb,amsmath,amsfonts,dsfont,hyphenat}
\usepackage{fixltx2e,mathptmx}
\usepackage[hyphens,T1]{url}
\usepackage[scaled=0.9]{helvet}
\usepackage{latexsym,tabularx,comment,colortbl,epstopdf,xspace}
\usepackage{times,multirow,moreverb,booktabs}
\usepackage{accents,tikz,xfrac,cleveref,xfrac}
\usepackage{paralist,fancyref,enumitem}
\usepackage[labelformat=simple]{subcaption}
\usepackage[scaled=0.925]{couriers}
\DeclareSymbolFont{cmbrightop}{OT1}{cmbr}{m}{n}
\newcolumntype{L}{>{\raggedright\arraybackslash}X}

\newcommand{\eat}[1]{}
\usepackage{color}

\definecolor{mycitegreen}{RGB}{0,90,0}
\definecolor{myrefblue}{RGB}{90,0,0}
\definecolor{red}{RGB}{140,0,0}
\definecolor{green}{RGB}{0,80,0}
\usepackage{multirow}
\usepackage{lipsum}
\PassOptionsToPackage{bookmarks=false}{hyperref}

\setitemize{noitemsep, topsep=0pt, parsep=0pt, partopsep=0pt,
leftmargin=1em}
\setenumerate{noitemsep, topsep=0pt, parsep=0pt, partopsep=0pt,
leftmargin=1.3em}

\DeclareRobustCommand\ttfamily
        {\not@math@alphabet\ttfamily\mathtt
         \fontfamily\ttdefault\selectfont\hyphenchar\font=-1\relax}
\DeclareTextFontCommand{\mytexttt}{\ttfamily\hyphenchar\font=45\relax}

\usepackage[figure,vlined]{algorithm2e}

\fancyrefchangeprefix{\fancyreffiglabelprefix}{f}
\fancyrefchangeprefix{\fancyreftablabelprefix}{t}
\fancyrefchangeprefix{\fancyrefeqlabelprefix}{eqn}
\renewcommand*{\figurename}{Figure}

\newcommand{\textred}[1]{\textcolor{red}{#1}}

\ifx\noeditingmarks\undefined
   \newcommand{\pgwrapper}[2]{\textred{#1: #2}}
\else
   \newcommand{\pgwrapper}[2]{}
\fi

\newsavebox{\savepar}

\let\oldbibliography\thebibliography
\renewcommand{\thebibliography}[1]{\oldbibliography{#1}%
  \setlength{\itemsep}{0pt}}

\SetKwFunction{Fn}{\textbf{Function}}%
\SetKwFunction{Dd}{D2Didle}%
\SetKwFunction{Nb}{NoBackoff}%
\SetKwFunction{Bk}{BackOff}%

\title{Data-Driven Path Selection for Real-Time \\ Video Streaming at the Network Edge}
\author{\large Sabur Baidya, Peyman Tehrani and Marco Levorato\\
\normalsize Donald Bren School of Information and Computer Sciences, UC Irvine\\
\normalsize e-mail: \{sbaidya,~peymant,~levorato\}@uci.edu\vspace{-0.5cm}}
\date{}

\usepackage[mathlines,switch]{lineno}
\usepackage{lipsum}
\makeatletter

\let\old@ps@headings\ps@headings
\let\old@ps@IEEEtitlepagestyle\ps@IEEEtitlepagestyle
\def\confheader#1{%
    \def\ps@IEEEtitlepagestyle{%
        \old@ps@IEEEtitlepagestyle%
        \def\@oddhead{\strut\hfill#1\hfill\strut}%
        \def\@evenhead{\strut\hfill#1\hfill\strut}%
    }%
    \ps@headings%
}
\makeatother

\confheader{%
        \parbox{\textwidth}{\centering This article has been accepted for publication in the IEEE International Conference on Communications (ICC) Workshop on ``Edge Machine Learning for 5G Mobile Networks and Beyond" 2020.}
}

\usepackage[pscoord]{eso-pic}
\newcommand{\placetextbox}[3]{
\setbox0=\hbox{#3}
\AddToShipoutPictureFG{ \put(\LenToUnit{#1\paperwidth},\LenToUnit{#2\paperheight}){\vtop{{\null}\makebox[0pt][c]{#3}}}
}
}
\placetextbox{.5}{0.065}{\footnotesize {\copyright 2020 IEEE. Personal use of this material is permitted. Permission from IEEE must be obtained for all other uses, in any current or future media} }
\placetextbox{.5}{0.055}{\footnotesize { including reprinting/republishing this material for advertising or promotional purposes, creating new  collective works, for resale}} 
\placetextbox{.5}{0.045}{\footnotesize { or redistribution to servers or lists, or reuse of any copyrighted component of this work in other works.}
}

\begin{document}

\maketitle
\vspace{1cm}

%\pagestyle{plain}
%\thispagestyle{plain}
%\pagestyle{empty}
%\thispagestyle{empty}

%%%%%%%%%%%%%%%%%%%%%%%%%%%%
%%%%%%    ABSTRACT    %%%%%%
%%%%%%%%%%%%%%%%%%%%%%%%%%%%
\begin{abstract}
\begin{comment}
The coexistence of a myriad of data streams and the topological characteristics 
of urban environments result into extreme variability of wireless links' capacity. 
The volatility of the wireless environment challenges the establishment of reliable distributed 
data processing and control loops, such as those 
characterizing edge computing systems supporting mission critical applications, at the network edge.
\end{comment}
In this paper, we present a framework for the dynamic selection of the 
wireless channels used to deliver information-rich data streams to edge servers. The approach we propose is data-driven, where a predictor, whose output informs the decision making of the channel selector, is built from available data on the transformation imposed by the network on previously transmitted packets.
The proposed technique is contextualized to real-time video streaming for immediate processing. 
The core of our framework is the notion of probes, that is, short bursts of packets transmitted over unused channels to acquire information while 
generating a controlled impact on other active links. Results indicate a high accuracy of the prediction output and a significant improvement in terms of received video quality when the prediction output is used to dynamically select the used channel for transmission.

%Alternate: Edge computing systems rely on reliable distributed data processing 
%and control loops from the wireless edge to the wireless servers. 
%However, wireless networks are less reliable because they vary over time and bandwidth   
%in the presence of interference from other nodes, congestion, and application-layer
%variations. We investigate a data-driven approach to make the wireless networks reliable 
%by predicting the future quality of the available links and dynamically selecting the wireless 
%link with the best quality-of-service. Our framework relies on the notion of probing 
%where client devices transmit short bursts of packets, called probes, to predict temporal characteristics
%of the unused links while minimizing the impact on the active data links. We evaluate our approach
%on real-world traces of video transmission over a network of wireless nodes and show significant performance 
%gains over open-loop approaches.
\end{abstract}

%%%%%%%%%%%%%%%%%%%%%%%%%%%%%%%%%%%%%%%%%%%%%%%%%%%%%%
%%%Introduction
%%%%%%%%%%%%%%%%%%%%%%%%%%%%%%%%%%%%%%%%%%%%%%%%%%%%%%

\vspace{-1mm}
\section{Introduction}
%\vspace{-0.3mm}
Recently introduced paradigms where machines collaborate to 
accomplish a common task are imposing increasingly stringent constraints 
on the Quality of Service (QoS) achieved by the underlying communication 
layers. An illustrious example of such model, which have been receiving considerable 
attention, is edge computing~\cite{bonomi2012fog,shi2016promise,baidya2017netselect}, where a compute-capable 
machine positioned at the network edge provides data processing services 
to mobile devices interconnected through a low-latency, one hop, wireless link.

However, the volatile nature of wireless links may induce periods of time 
where the ability of mobile devices to deliver data to edge servers degrades. 
For instance, bursts of data transmissions from other wireless terminals active 
on the same frequency channel, or temporary loss of line of sight connectivity to the access point or 
base station in urban environments may drastically reduce achievable data transfer rate and result in large delays or packet loss. Intuitively, these issues are particularly relevant when edge computing is used to support real-time mission critical applications.

The overall objective of this paper is to maximize the integrity of data streams emitted from mobile devices and directed to the network edge under stringent capture-to-delivery delay constraints imposed by real-time applications. We specifically focus on one of the most challenging and relevant applications: real-time video streaming for immediate edge-assisted analysis. We
consider a scenario where a mobile device and an edge server can communicate over multiple wireless channels -- for instance, multiple Wi-Fi channels associated with one or more access points, and make the following assumptions:
(\emph{i}) due to mobility and activity from other active devices the characteristics -- \emph{e.g.}, short term throughput and packet-wise delay -- of each channel present complex dynamics at different time scales; and (\emph{ii}) devices can acquire information on the channel state -- abstracted here as a \emph{transformation} imposed to the packet stream -- only by transmitting packets over available channels and observe their outcome.
In this challenging environment, we formulate a channel selection problem whose goal is twofold: (\emph{i}) maximize the perceived quality of the video stream at the edge server, while (\emph{ii}) minimizing the impact of the information flow on the Quality of Service (QoS) of other active data streams. At the core of the problem is reliable channel \emph{prediction}, that is, the ability of the channel selection logic to forecast future performance of the video stream to inform selection decisions. 
Critically, as only the channel currently being used is \emph{observed}, the selection logic is clueless about the current state of the other channels. In order to acquire complete state information, all the channels would need to be used simultaneously, thus %considerably increasing network load,
possibly degrading the QoS of other data streams. To solve this impasse, we propose a \emph{probe-based} approach, where prediction over unused channels is informed by the transformation applied by the channel to a train of short packets periodically transmitted.

In summary, the paper makes the following contributions:

\vspace{1mm}
\noindent
(\emph{a}) By detailed simulation, we create a comprehensive datasets based on real-world application layer traffic traces.

\vspace{1mm}
\noindent
(\emph{b}) We characterize the information on the future Peak Signal to Noise
Ratio (PSNR) of the video stream contained in features of the packet streams -- both video and probes.

\vspace{1mm}
\noindent
(\emph{c}) We develop a prediction framework based on state of the art classification algorithms, including Convolutional Neural Networks (CNNs).

\vspace{1mm}
\noindent
(\emph{d}) Based on the predictor's output, we implement an algorithm dynamically selecting the channel which most likely will provide the highest PSNRs in a predefined temporal window.

\vspace{1mm}
The rest of the paper is organized as follows. Section~\ref{sec:relwork} discusses related work. Section~\ref{sec:preliminaries} presents the considered scenario and channel selection problem, and introduces the prediction strategy. In Section~\ref{sec:dataset}, we describe the network environment, and the dataset.% used 
%to build and test the channel selection 
%for the framework.
Section~\ref{sec:prediction} presents the classification and selection approach.
In Section~\ref{sec:eval}, we provide
performance evaluation and
%extensive results on the performance of offline classification and online channel selection. Finally,
Section~\ref{sec:concl} concludes the paper.

\section{Related Work}
\label{sec:relwork}
%\vspace{-0.8mm}
Real-time video streaming over wireless networks often needs to compromise its QoS to adapt to the inconsistencies in the communication channels. Dynamic Adaptive Streaming over HTTP (DASH) \cite{stockhammer2011dynamic} and similar protocols \cite{kua2017survey, zambelli2009iis} were developed to adapt the quality of the transmitted video -- that is, its resolution, frame rate, etc. -- to the capacity of the channel. However, leveraging other available channels can improve the QoS of the video streaming. Protocols, e.g., Multipath TCP (MPTCP) \cite{barre2011multipath} was developed to harness the capacity of multiple communication paths, however, introduce challenges in terms of packet reordering and head-of-line blocking in case of high diversity among the paths. Based on MPTCP, Multi-Path DASH (MPDASH) \cite{han2016mp} has been proposed to extend DASH to multi-path scenarios. However, MPDASH is based on MPTCP, and thus comes with all the limitations of TCP in a real-time streaming scenarios. Additionally, 
TCP-based techniques are not a good match to wireless scenarios due to their inherent fast variability. Finally, these techniques presume the use of all the channels simultaneously, with a possibly long adaptation time when a new path is added or removed, an event which may frequently occur in mobile wireless networks. Due to all the aforementioned reasons, for real-time video streaming, we choose to use the best of the available communication paths and use streaming over UDP.

However, proactively selecting the best wireless path ensuring stability of the video over a time window needs accurate prediction of the wireless network conditions. 
A number of recent contributions use machine learning techniques to predict the state of wireless channels.
For instance, a decision tree based classification algorithm  is employed in \cite{abbas2013learning} to choose the best available network between a 3G and a WLAN using features such as RSSI, location and type of application. In \cite{kajita2015throughput}, Support Vector Machines (SVM) is used to estimate the throughput and delay 
%of an Access Point's user 
in a WiFi network, given the traffic volume and signal strength of active interferes  as input features.
A recent work presents ``PenSeive'' \cite{mao2017neural}, an Adaptive Bit Rate (ABR) video transmission framework based on reinforcement learning and neural networks. The primary goal is to learn from past decision on output bit rate with respect to channel state. However, these prediction techniques are focused on the physical layer, with no consideration of channel use patterns. Here we address a problem where due to the presence of other data streams, only Physical layer based frameworks, e.g., pure SNR-based prediction fail to capture the interaction of upper layer complexities for the coexistence.

%\input{motivation.tex}

%%%%%%%%%%%%%%%%%%%%%%%%%%%%%%%%%%%%%%%%%%%%%%%%%%%%%%
%%%Discussion
%%%%%%%%%%%%%%%%%%%%%%%%%%%%%%%%%%%%%%%%%%%%%%%%%%%%%%

\section{Preliminaries}
\label{sec:preliminaries}

%From a high-level perspective, 
The development of a predictive framework to  dynamically select channels providing the desired quality to a data stream in presence of complex underlying traffic and channel gain presents two main technical challenges:

(\emph{ii}) Develop an effective methodology to acquire information on the state of all available channels; and

(ii) Build effective strategies mapping the acquired state information onto future QoS or application layer performance of the data stream.

\begin{figure}[!t] 
\centering
        \includegraphics[width=0.9\columnwidth]{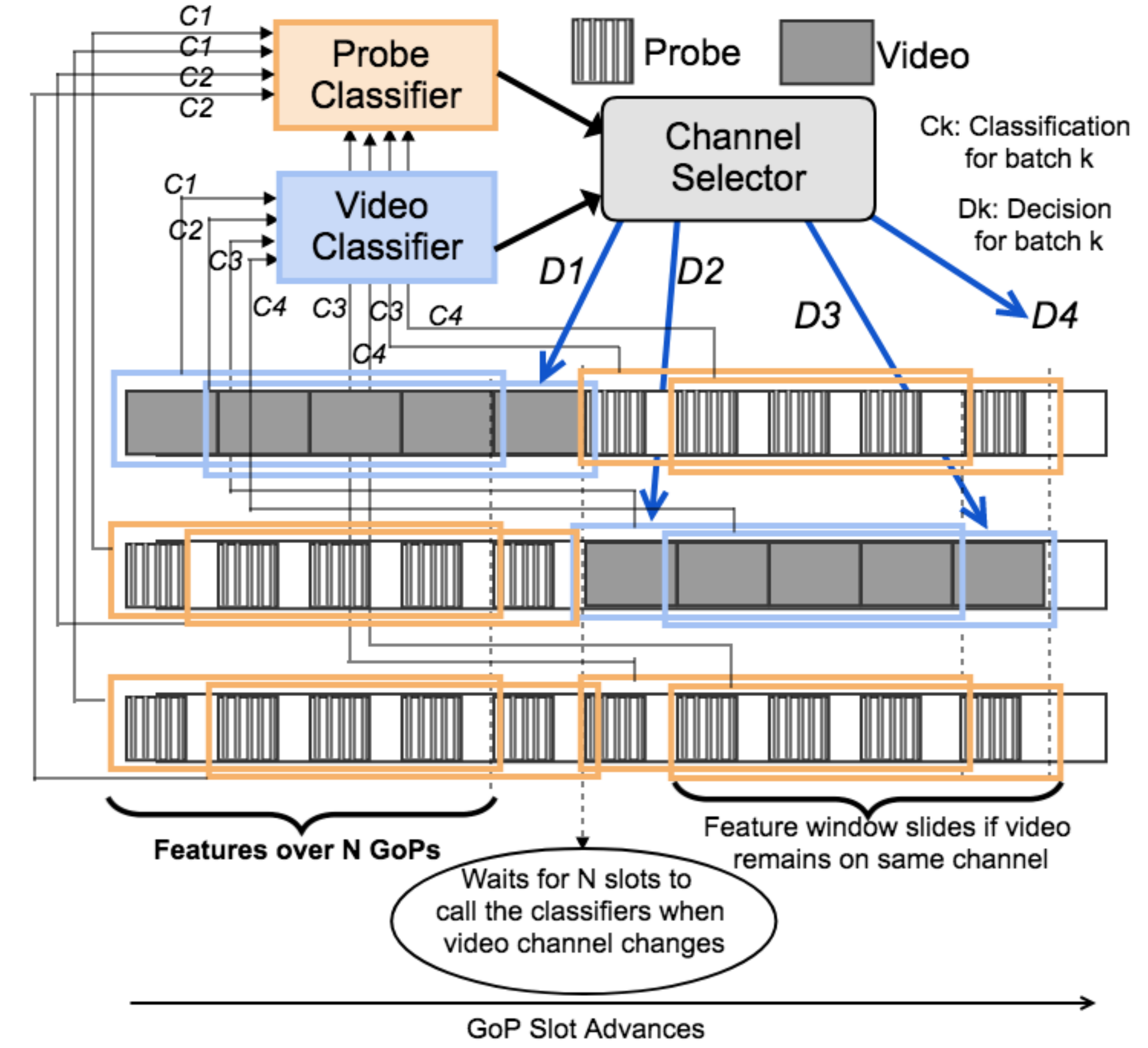}
        \caption{High level schematics of the channel selection framework based on probe and data classifiers.}
        \label{fig:scheme}
        \vspace{-4mm}
\end{figure}

Fig.~\ref{fig:scheme} shows the general diagram of the proposed framework: features extracted from the video and probe streams are used to predict the future quality of the video if any of the $C$ channels were to be used for transmission. The individual components of the framework and the rationale behind its design are explained in the following.

\subsection{Scenario}

We consider a scenario where a video stream captured in real-time by a mobile user needs to be transported to an edge server. The user and the edge server can communicate through $C$ independent wireless channels -- for instance corresponding to different access points in the vicinity -- indexed with $c{=}1,\ldots,C$. The channels are shared with other users and data streams, which all contend for available frequency resources. 

The video stream is encoded, packetized using UDP, and transmitted over one of the wireless links to the edge server. The frame rate of the video is $F_r$~frames/s, and we assume that all the packets referring to frame $i$, $i{=}0,1,2,\ldots$ are injected into the transport layer socket at time $i/F_r$.

\subsection{Selection Problem}

Due to practical considerations, we define the selection process at the temporal granularity of Group of Pictures (GoPs). Let's index GoPs with $p{=}1,2,\ldots$. We, then, denote the average Peak Signal-to-Noise Ratio (PSNR) of GoP $p$ with $\Phi(p)$, computed as the average PSNR of all the component frames, and define the control variable $c(p){\in}(1,\ldots,C)$ corresponding to the channel selected for the transmission of GoP $p$.

In order to maximize the quality of the video, we simply aim at the selection of the channel providing the maximum average PSNR, that is,
\begin{equation}
    c(p){=}\arg\max_{c} \Phi(p,c).
\end{equation}
Rather clearly, the selector needs to make the decision before the GoP is transmitted, and selection is necessarily based on prediction.

\subsection{Prediction}

 We define prediction as a classification problem, where classes are defined based on functionals of the GoPs' PSNR within a window of future GoPs. This is motivated by the inherent difficulty in the short-term prediction of individual GoPs' PSNR values in network environments characterized by mobility and complex traffic patterns. Let's set the current GoP index to $p{=}0$, we then define $Z$ classes $z{\in}(1,\ldots,Z)$ as
\begin{equation}
    z = \mathcal{F}_W(\boldsymbol{\Phi}_{1,W}),
\end{equation}
where $\mathcal{F}_W$ is a functional mapping of the vector $\boldsymbol{\Phi}_{1,W}{=}(\Phi(1),\ldots,\Phi(W))$ to the class $z$.

Several choices of the classification function $\mathcal{F}_W$ are possible, including a simple discretized average over the PSNRs within the temporal window. Herein, we choose a function which aims at preserving a general good quality of most GoPs in the window. Specifically, we define
\begin{equation*}
    \mathcal{F}_W(\boldsymbol{\Phi}_{1,W}){=}\begin{cases} 1 & {\rm if} ~~\sum_{p{=}1}^W \mathit{I}(\boldsymbol{\Phi}(p)>y){\geq}k,\\
    0 & {\rm otherwise},
    \end{cases}
\end{equation*}
where $y$ is a predefined PSNR threshold, $k{\in}\{0,\ldots,W\}$ and $\mathit{I}(\cdot)$ is the indicator function. Thus, we say that the channel in the window is ``\emph{good}'' ($z{=}1$) if at least $k$ GoPs have a PSNR larger than $y$.
The selector, then, will attempt to assign the stream to a \emph{good} channel.

We now define the input of the predictor. As mentioned earlier, due to both technological and environmental parameters and their variations, the channel applies a transformation to the packet transmitted by the mobile user. Define  $\boldsymbol{\Omega}(p)=\left[\begin{matrix}\boldsymbol{\tau}(p)\\\boldsymbol{\lambda(p)}\end{matrix} \right]$ as the matrix composed of the row vectors $\boldsymbol{\tau}(p)$ and $\boldsymbol{\lambda}(p)$, whose elements are the injection time and size of all the packets transmitted during GoP $p$. We characterize such transformation using the function $\mathcal{T}(\cdot)$ as follows
\begin{equation}
    \left[\begin{matrix}\tilde{\boldsymbol{\tau}}(p)\\\tilde{\boldsymbol{\gamma}}(p)\end{matrix}\right]=\mathcal{T}(\boldsymbol{\Omega}(p),{\bf s}(p)),
\end{equation}
which takes as input $\boldsymbol{\Omega}(p)$ and a generic vector of latent variables ${\bf s}(p)$ to return the vectors $\tilde{\boldsymbol{\tau}}(p)$ and $\tilde{\boldsymbol{\gamma}}(p)$, which contain the delivery time and a delivery flag (equal to $1$ if the packet is delivered by the deadline and $0$ otherwise), respectively.

We, then, define the predictor $\overline{z}=\mathcal{P}(\boldsymbol{\Pi})$, where $\overline{z}{\in}\{0,1\}$ is the predicted class and $\boldsymbol{\Pi}$ is a matrix
\begin{equation}
    \boldsymbol{\Pi}=\left[\begin{matrix} f_1(0) & f_1(-1) & \ldots & f_1(-H) \\
    \vdots&\vdots&\ldots&\vdots\\
    f_F(0) & f_F(-1) & \ldots & f_F(-H)\end{matrix} \right]
\end{equation}
whose rows correspond to different features and columns to different GoP slots. The features are extracted from the matrix
\begin{equation}
    \left[\begin{matrix} \tilde{\boldsymbol{\tau}}(0)& \tilde{\boldsymbol{\tau}}(-1)&\ldots&\tilde{\boldsymbol{\tau}}(-H)\\\tilde{\boldsymbol{\gamma}}(0)&\tilde{\boldsymbol{\gamma}}(-1)&\ldots&\tilde{\boldsymbol{\gamma}}(-H) \end{matrix}\right].
\end{equation}

We note that the features needs to be calculated at the access point or edge server, which feeds them back to the user by means of a short packet. In order to ensure a reliable delivery of this packet, features associated with the last frame of the last GoP may be omitted.

\subsection{Probing}

 The input of the predictor are features computed from the outcome of packets transmitted over the channel in the $T$ preceding temporal slots (corresponding to GoPs). Clearly, computing such features is possible only if the channel currently used to transmit the video. To overcome this issue, we introduce and study in this scenario the notion of \emph{probes}. The core idea is to acquire information by periodically transmitting short trains of packets. This reduces the impact on other active streams compared to transmitting sections of the video.

%\input{Prob_formulation.tex}

%%%%%%%%%%%%%%%%%%%%%%%%%%%%%%%%%%%%%%%%%%%%%%%%%%%%%%
%%%Dataset
%%%%%%%%%%%%%%%%%%%%%%%%%%%%%%%%%%%%%%%%%%%%%%%%%%%%%%

\section{Network Environment and Dataset}
\label{sec:dataset}

Here we focus on the Wi-Fi communication technology, which has a high degree of unpredictability due to the contention mechanism and the low amount of control imposed by the infrastructure, and associate each of the $C$ channels to an access point. The investigation of a multi-technology environment, for instance including connection to LTE eNodeBs is left to future studies. 
We consider the following coexisting applications: (\emph{1}) Youtube video streaming, (\emph{2}) Skype voice call, (\emph{3}) File Transfer Protocol, (\emph{4}) Skype video call and (\emph{5}) collection of background web traffic that persistently inject traffic in the network. 

\begin{figure}[!t]
    \centering
    \includegraphics[width=0.85\columnwidth]{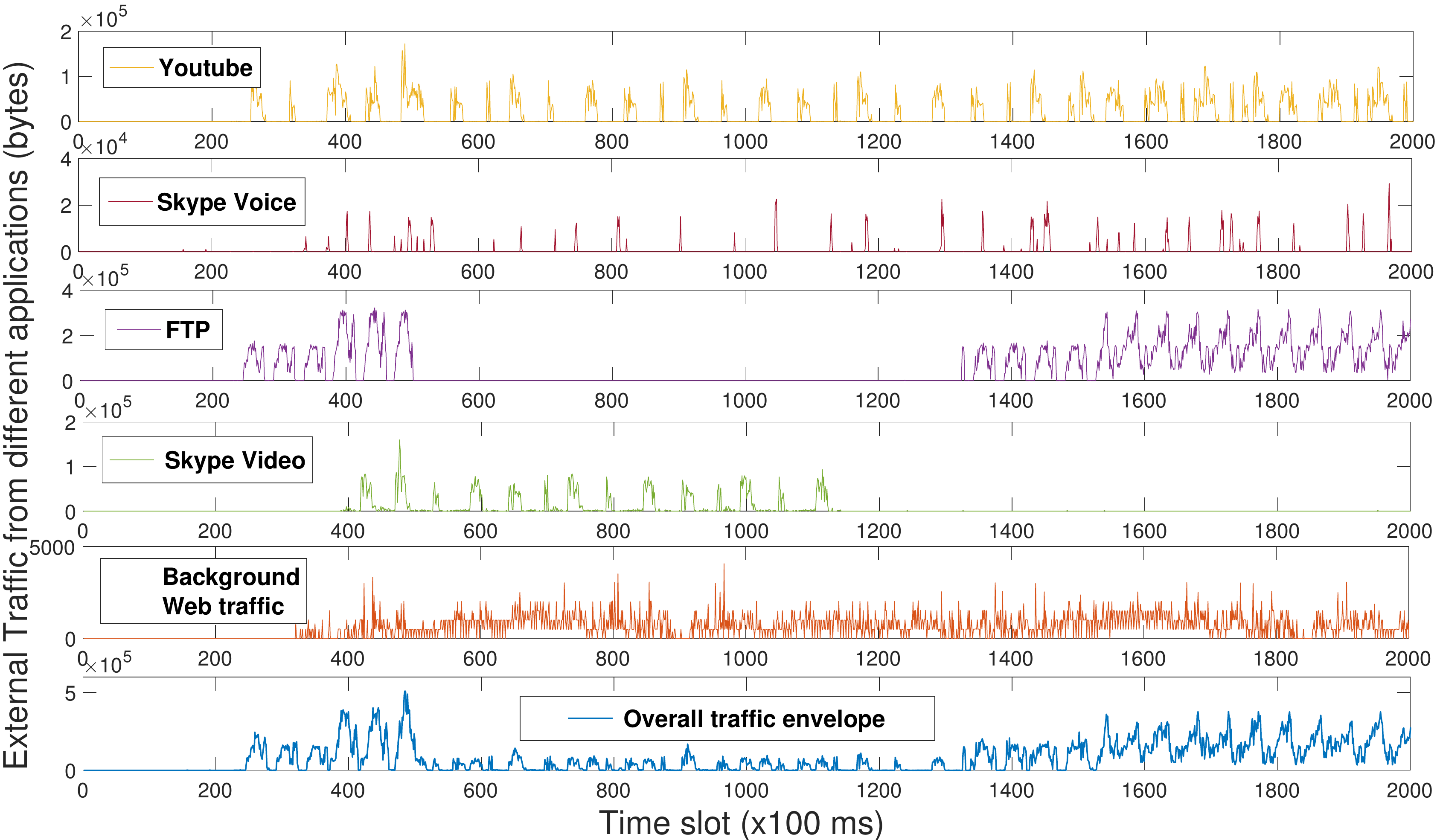}
    %\vspace{1mm}
    \caption{Envelope of the traffic classes at the network level. The set includes Youtube, Skype voice and video, and FTP applications as well as persistent background web traffic.}
    \vspace{-4mm}
    \label{fig:envel}
\end{figure}

Figure~\ref{fig:envel} shows the traces collected for the mentioned classes of applications and combined complex structure of the overall envelope of the traffic shape when all of them are active. We inject these applications with stochastic activation/deactivation patterns.

The used video stream has a resolution of 1920x1080 and encoded with H.264/AVC~\cite{wiegand2003overview} format with average encoding bitrate of $1 Mbps$ using ffmpeg~\cite{tomar2006converting}.
 We generate a series of Transport Stream (TS) packets and transmitted over UDP.
 %of fixed size $188$ bytes and use UDP as transport layer. Several TS packets are encapsulated in UDP packet and transmitted over the network. 
 All the other coexisting data streams are transported over TCP.
 
 To extract the dataset, we create $3$ non-overlapping WiFi channels in the NS-3 simulator, and run simulations for $50000$ GoPs. Each GoP is composed of $30$ frames transmitted over $1$ sec. Each video frame of the GoP is transmitted as a burst of packets. Since the video is temporally encoded, the first frame of the GoP is a reference frame with larger size of $42$ UDP packets of $1400$ bytes, all following differential frames are smaller and contained in a burst of $2$ to $4$ UDP packets. Within any GoP slot, we transmit the video data on one of the $3$ channels and a burst of probe packets at the beginning of the GoP on the the other available channels (see Fig.~\ref{fig:scheme}).
 
\begin{figure}[tb]
\centering
        \includegraphics[width=0.8\columnwidth, height=0.4\columnwidth]{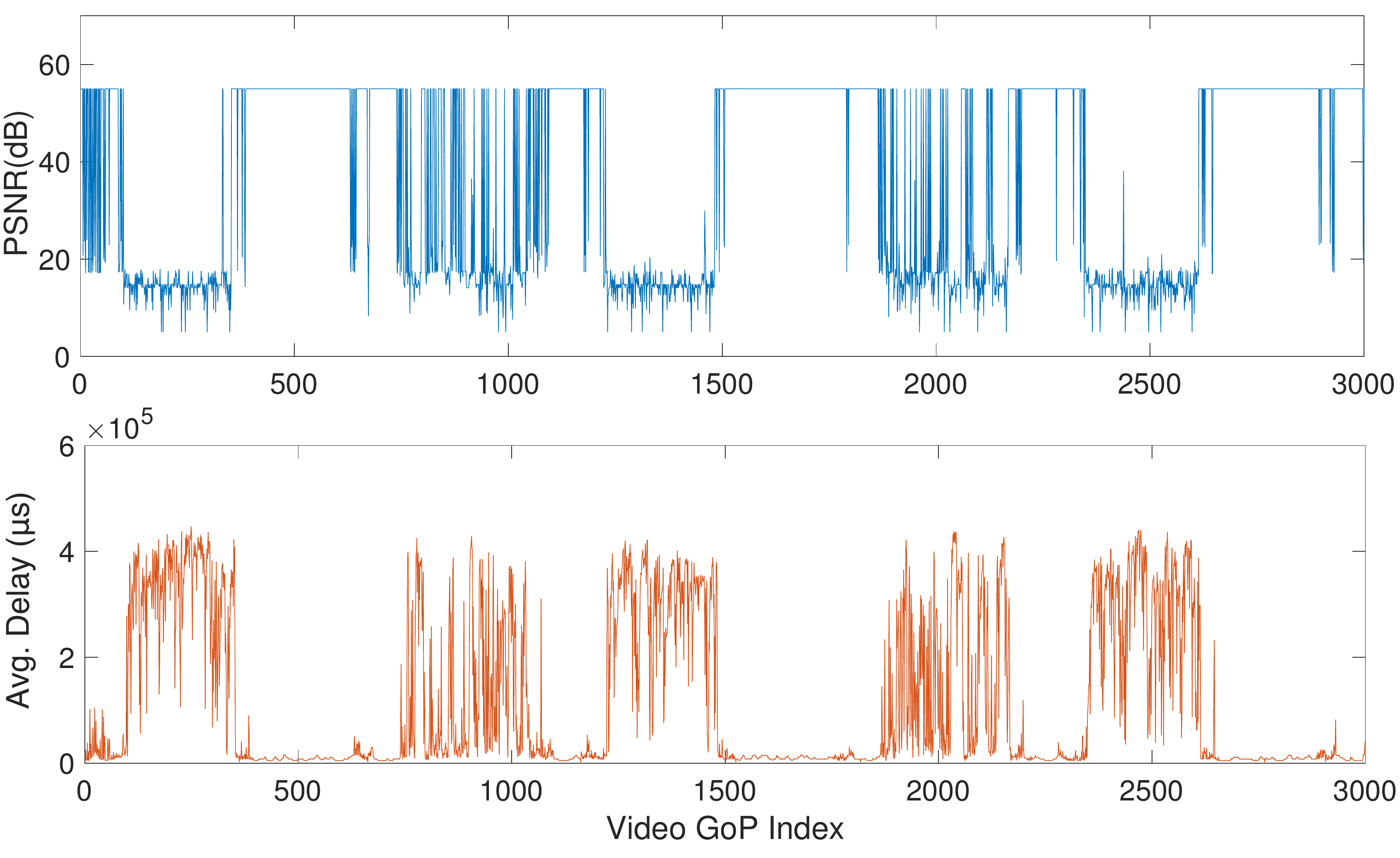}
        \caption{Temporal pattern of PSNR and average packet delay induced by the traffic traces and their activation/deactivation.
        }
        \vspace{-6mm}
        \label{fig:traces}
\end{figure}

 Fig.~\ref{fig:traces} depicts an example of temporal trace of the PSNR and average packet delay at the GoP granularity. Periods of high channel load with different characteristics can be observed, interluded by periods of milder traffic conditions. Note the non-trivial relationship between packet delay and PSNR due to the complex characteristics of video encoding.

%\input{dataset.tex}

%%%%%%%%%%%%%%%%%%%%%%%%%%%%%%%%%%%%%%%%%%%%%%%%%%%%%%
%%%Dataset
%%%%%%%%%%%%%%%%%%%%%%%%%%%%%%%%%%%%%%%%%%%%%%%%%%%%%%

\section{Prediction Framework}
\label{sec:prediction}

The dataset described in the previous section is used to build the predictor $\mathcal{F}$. 
Note that we need two distinct classifiers, translating video stream and probe stream features into the binary class we use to represent the quality of future video GoPs.
%Note that two separate prediction functions are used for probe-based and video-based prediction
%\textcolor{blue}
{We compute the features over a
temporal window −H,...,0, which we refer to as the history.
To define the classes z=1 and 0 – high and low video quality,
respectively, we set the PSNR threshold to 40 dB, and the number
of above-threshold GoPs to k=6. As in the dataset collection,
we set the prediction and history window size to 10.}

\begin{figure}[tb]
\centering
        \includegraphics[width=0.9\columnwidth, , height=0.65\columnwidth]{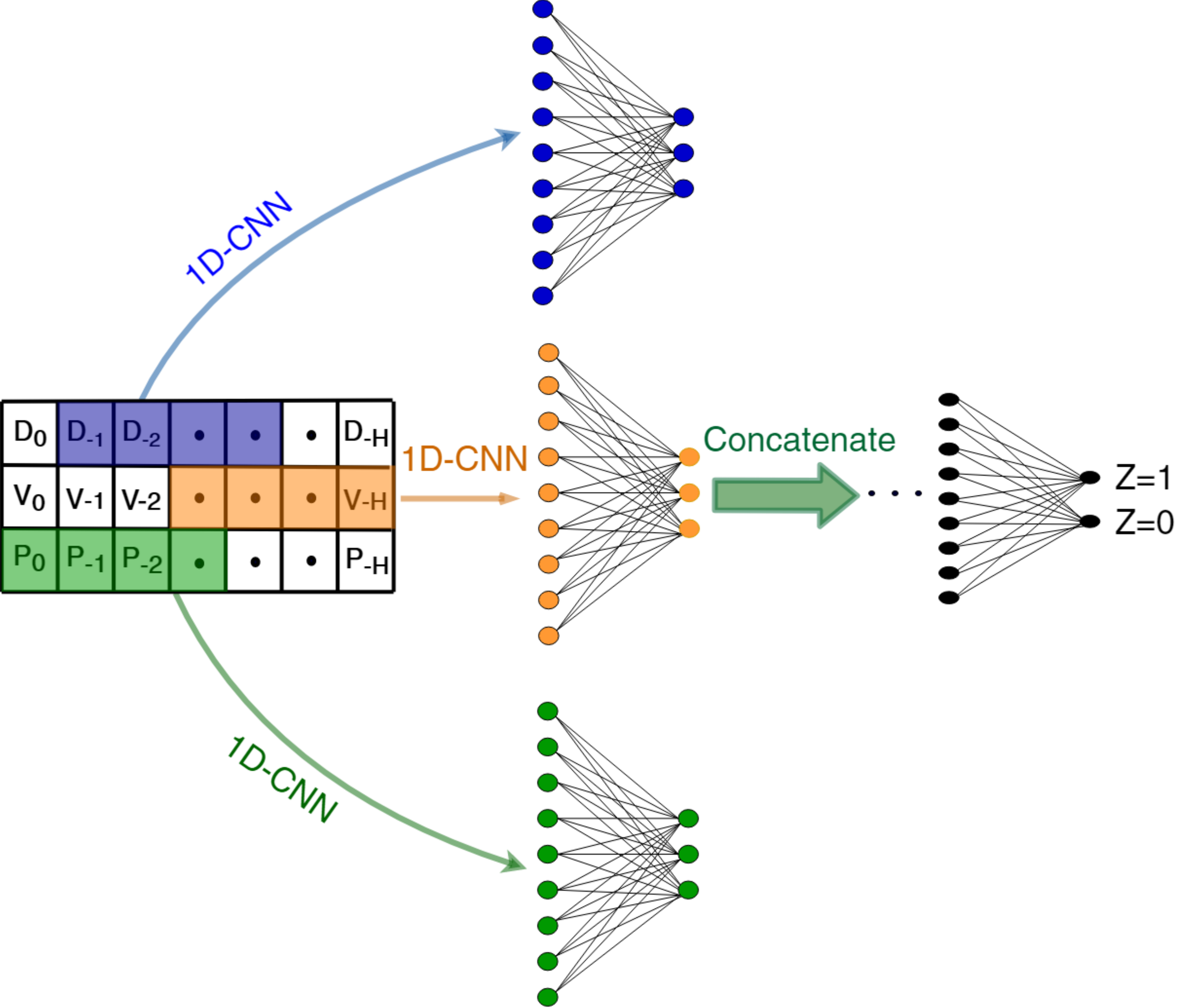}
        \caption{CNN Architecture used to build the predictor. The model takes as input a vector with shape $3 \times (H+1)$, in which each row is associated with one feature type $D$,  $V$ and $P$ to indicate average delay, delay variance and packet loss, respectively. Then a 1D CNN is applied to each row to analyze the temporal fluctuations and trends of each feature type, and a the filtered data is unrolled and fed to a fully connected neural network to extract the output class label.}
        \vspace{-4mm}
        \label{fig:CNN Art}
\end{figure}

In order to fully exploit the temporal dependencies which exist between the features and the future class, we use a deep Convolutional Neural Network (CNN)~\cite{lecun1998gradient}. This choice also allows the use of the CNN soft labels' output to improve selection accuracy, as explained later. The architecture of CNNs is designed to take advantage of the multi-dimensional structure of the input. This is achieved using local connections and weights followed by pooling, which results in translation invariant features.
In the considered case, we use the feature matrix $\boldsymbol{\Pi}$ as a 2-D input set feature. Then, $\Lambda$ different 1-D kernels (filters) are applied on the each row of the input matrix. Assuming that the size of all kernels is the same, and is equal to $T_k$ which $T_k<H$. Kernel ${K}_{\lambda}$ can be described as a vector as below:
\begin{equation*}
\vspace{-1mm}
{K}_{\lambda} = \begin{bmatrix}
       \kappa_{0}^{\lambda} & \kappa_{2}^{\lambda} & ... & \kappa_{T_k-1}^{\lambda}
     \end{bmatrix}
%\vspace{-1mm}
\end{equation*}
The output of the convolution between ${K}_{\lambda}$ and input feature vector is a vector with length $L=H-T_k+1$ in which the element $R_{\lambda}^{i}(l)$ is obtained as follows:
\vspace{-2mm}     
\begin{equation}
\label{eq:conv}
R_{\lambda}^{i}(l)=\sum_{p=0}^{T_k-1}\kappa_{p}^{\lambda}f_i(l-p)
\end{equation}
Where $f_i$ in the equation above can be any of three feature $f_1$, $f_2$ or $f_3$.
After convolving with the features, each kernel roughly summarizes key trends in the feature progression. For example, a kernel can detect whether there is a decreasing or increasing trend in recent slots, or capture a peak or valley within the window. A larger number of kernels allows the model to capture a broader range of fluctuations and trends in the features' history. 
After that, all $R_{\lambda}^{i}(l)$ are unrolled and passed through a non linear activation function,  which is, then, fed to a deep fully connected neural network as shown in Fig. \ref{fig:CNN Art} to output the class labels.

\subsection{Channel Selection}

The goal of selection decisions is to maximize the short-term future PSNR of the video given the current state of all the channels. Due to prediction errors, a channel whose predicted label is $z{=}1$ may still present a larger than expected number of low PSNR GoPs. To mitigate the impact of prediction errors, we \emph{rank} the channels using the soft output of the corresponding predictors and select the channel with the highest rank. Therefore, the policy for the channel selector is to choose a $c^{*}$ channel that
\begin{equation}
\vspace{-1mm}
    c^{*}{=}\arg\max_{c} \hat{z_c},
\end{equation}
where $\hat{z}_c$ is the soft output of the classifier in channel $c$.

%\input{Classification.tex}

%%%%%%%%%%%%%%%%%%%%%%%%%%%%%%%%%%%%%%%%%%%%%%%%%%%%%%
%%%Discussion
%%%%%%%%%%%%%%%%%%%%%%%%%%%%%%%%%%%%%%%%%%%%%%%%%%%%%%
%\input{discussion.tex}

%%%%%%%%%%%%%%%%%%%%%%%%%%%%%%%%%%%%%%%%%%%%%%%%%%%%%%
%% System Description
%%%%%%%%%%%%%%%%%%%%%%%%%%%%%%%%%%%%%%%%%%%%%%%%%%%%%%
%\input{system.tex}

%%%%%%%%%%%%%%%%%%%%%%%%%%%%%%%%%%%%%%%%%%%%%%%%%%%%%%
%% Implementation and Evaluation Results
%%%%%%%%%%%%%%%%%%%%%%%%%%%%%%%%%%%%%%%%%%%%%%%%%%%%%%

\section{Performance Evaluation}
\label{sec:eval}

We provide here an extensive evaluation of the offline and online performance of classification and channel selection. We used PyTorch~\cite{ketkar2017introduction} to build the CNN model, and the trained model is integrated into the NS-3 network environment for online performance test. To build the CNN model, we use $\Lambda=10$ different kernels and set the kernel length to $T_k=5$ which is equal to half of the history size. We set the learning rate to $l_r=0.003$ and use the Adam optimizer \cite{kingma2014adam} to train the model. We train and test the classifier on the probe and video datasets. Each dataset includes approximately $10000$ video and $2000$ probe datapoints. We use $\%75$ of datapoints for training and the remaining $\%25$ to assess offline classification performance. Our model takes about 5 microseconds for inference and is pre-loaded for parallel execution in another processing unit, e.g. GPU at the edge. The inference time (5 microsec) being negligible compared to the GoP duration (1 sec) does not affect the feature values which we compute based on  packet transmission in one GoP. So, we reduce the packet arrival deadlines from 1 sec to (1-0.000005) = 0.999995 sec to maintain real-time prediction and channel selection.   

\begin{figure}[t]
\begin{center}
        \includegraphics[width=0.8\columnwidth, height=0.5\columnwidth]{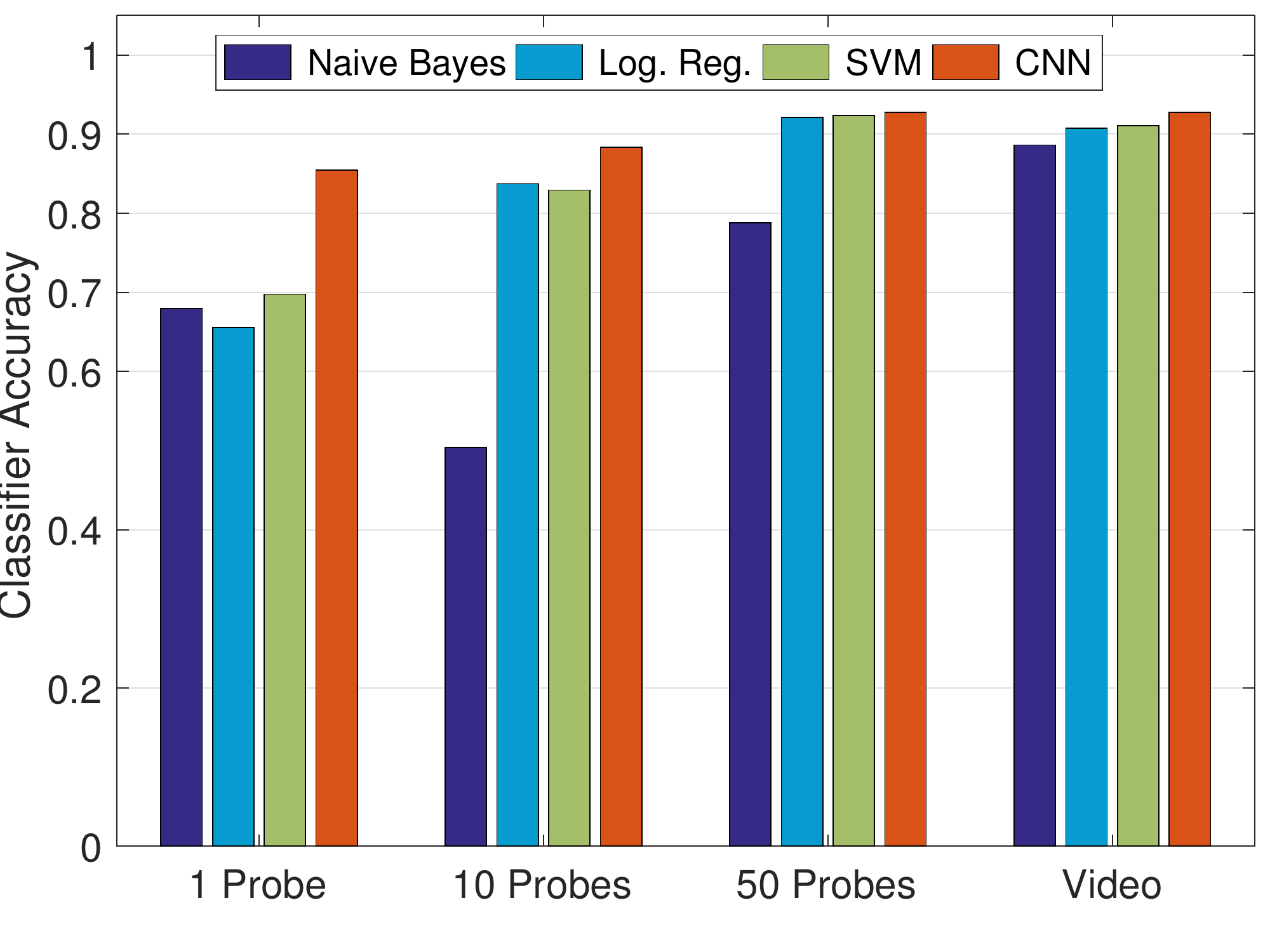}
        \caption{Performance Comparison of different classifiers for different batch sizes of 100 bytes probe and the video stream-based classifier.}
        \label{fig:compare}
        \vspace{-5mm}
\end{center}
\end{figure}

\begin{figure}[t]
\begin{center}
        \includegraphics[width=0.8\columnwidth,  height=0.5\columnwidth]{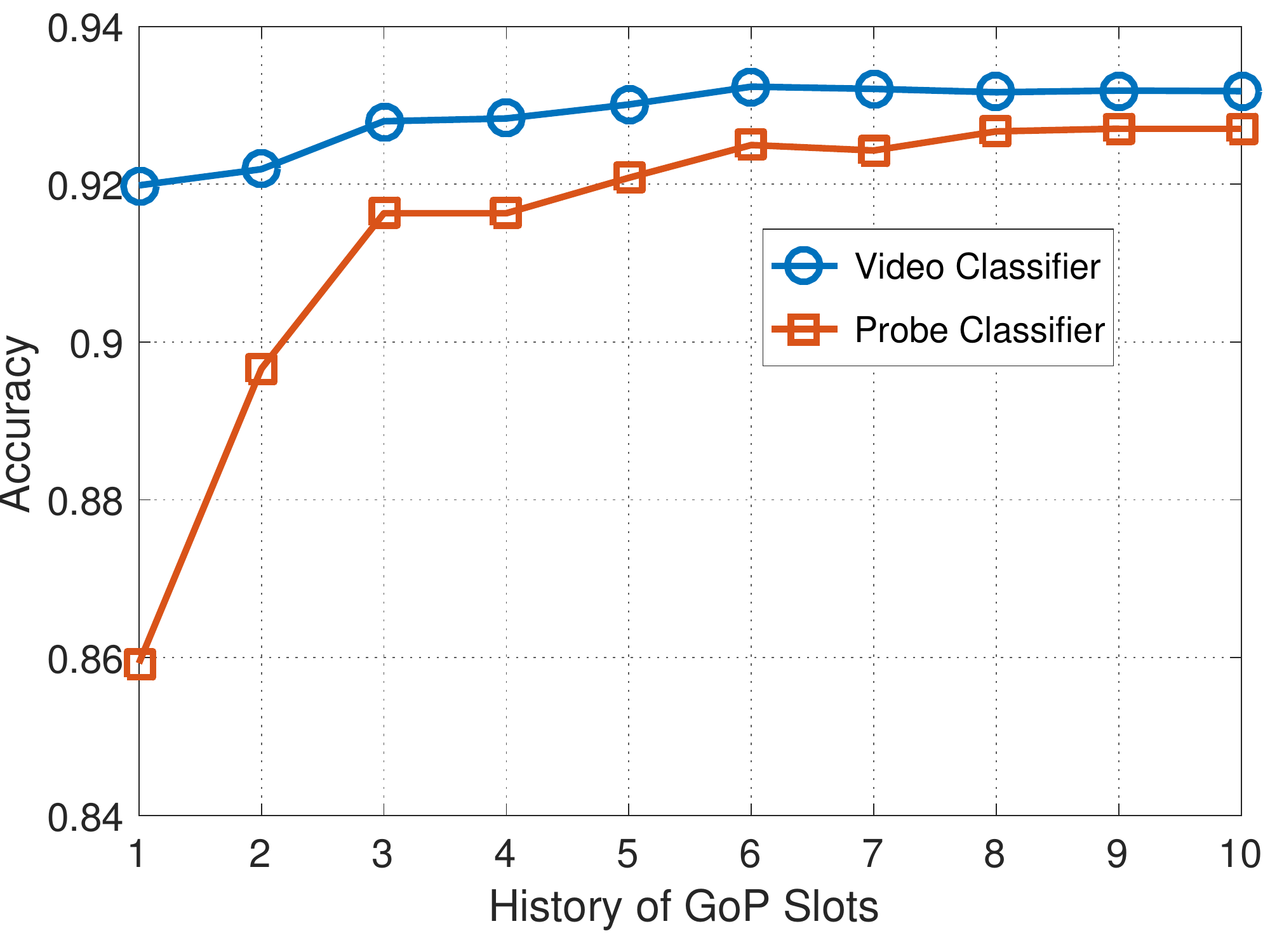}
        \caption{Comparison of the effect of the history length on classification accuracy with CNN based classifier.}
        \vspace{-8mm}
        \label{fig:comphist}
\end{center}
\end{figure}

\subsection{Offline Classifier Performance}

We first report micro-benchmarks on the performance of the classifier model in terms of
prediction accuracy. 
Figure~\ref{fig:compare} compares the classification accuracy achieved by probe composed of a different number of packets (packet size equal to $100$~bytes) and that achieved based on the video stream. Interestingly, the large amount of information collected by past video packets -- which are larger in size and number compared to probes -- allows accurate classification even when using simple classifiers. Conversely, accurate classification based on lightweight probes necessitates complex classifiers, capable of separating the two classes using more convoluted surfaces. This motivated our choice of CNN as the core of the proposed channel selection framework. 

We analyze the effect of the history size $H$ on the classification performance in Fig.~\ref{fig:comphist}. Interestingly, while the classifier based on the video packet stream suffers a relatively small accuracy loss when the history size is reduced, the probe-based classifier necessitates a larger amount of GoP slots to provide comparable performance. This is connected to the need of a larger set of packets spread over time to adequately sample the network state and its dynamics. 

\subsection{Online Channel Selection}

\begin{figure}[!t]
\begin{center}
        \includegraphics[width=0.8\columnwidth]{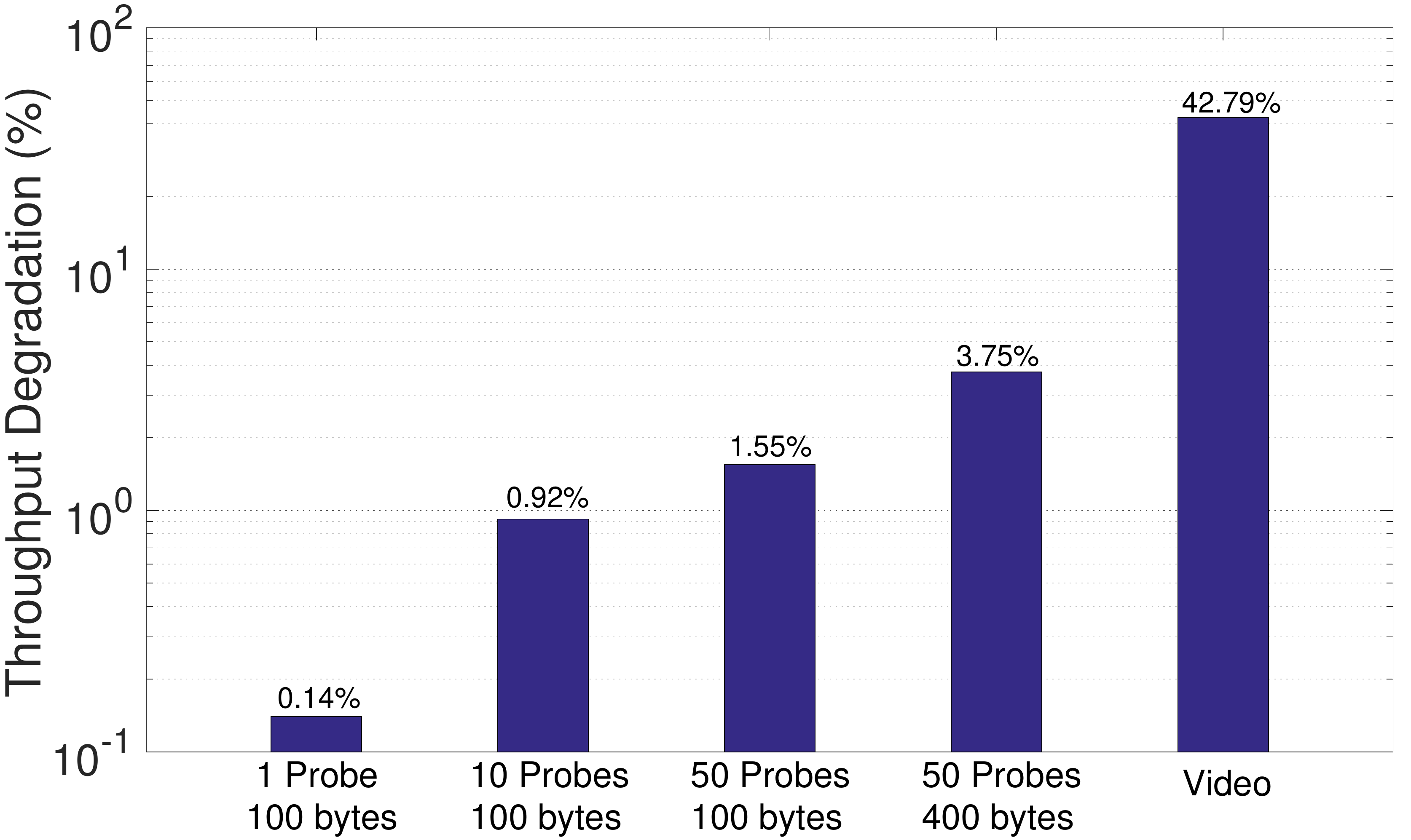}
        \caption{Throughput degradation of other applications imposed by probes or video streams.}
        \vspace{-6mm}
        \label{fig:thrpt_degrad}
\end{center}
\end{figure}

First, in order to fully motivate the use of probes, we measure the degradation -- in percentage with respect to the case with no video or probe stream -- to other data streams' throughput caused by different probe models and the video stream itself (Fig.~\ref{fig:thrpt_degrad}). We can observe that an already effective set of probes (burst size equal to $10$) causes a $1$\% throughput loss, compared to the $43$\% caused by the video stream.

We now analyze the performance of the data-driven channel selection framework in terms of output video quality.
Based on our offline evaluation of the classifiers in Fig ~\ref{fig:compare} and~\ref{fig:comphist}, here we use 50 probe packets for probe classifier and use a history of 10 GoP history slots for the features to achieve the  accuracy comparable to the video classifier.

 \begin{table}[!t]
\begin{tabular}{p{8cm}p{16cm}}
\small
\vspace{1mm}
\begin{tabular}{c|c|c}
%\centering
\hline
Avg. Encoding Bitrate & FileSize (Bytes) & Avg. PSNR (dB)\\
\hline
800 Kbps & 104977 & 36.95\\
600 Kbps & 78772 & 35.82\\
400 Kbps & 53241 & 35.01\\
200 Kbps & 27486 & 30.03\\
100 Kbps & 16056 & 26.65\\
50 Kbps & 13912 & 24.47\\
25 Kbps & 10865 & 18.63\\
\hline
\end{tabular}     
\end{tabular}
% \vspace{2mm}
\caption{Mapping of bitrate to file size and achievable PSNR for ABR based streaming of the selected video.}
\vspace{-4mm}
\label{table:abr}
\end{table}

We compare the data-driven channel selector with the following cases:

\vspace{1.5mm}
\noindent
{\bf (i)} \emph{Fixed Channel:} The entire video is sent over one fixed channel which gives the lower bound of the performance.

\vspace{1.5mm}
\noindent
{\bf (ii)} \emph{Adaptive Bit Rate (ABR) with full knowledge:} Assuming full knowledge of the channel throughput, we implement an ABR framework matching the available throughput to a coding rate. In order to compute the PSNR, we do not adapt the resolution of the video as in DASH~\cite{stockhammer2011dynamic} but adapt the encoding bitrate. Table~\ref{table:abr} shows how different encoding bitrate maps the size of the video segment with corresponding achievable PSNR. 
%We, then, pre-encode different versions of the video at different bitrates and use the map reported in table~\ref{table:abr} for online adaptation. 
In realistic conditions, the adaptation of the encoding rate in real-time streams is rather difficult due to the high computation demand.

\noindent
{\bf (iii)} \emph{Delay-Based Simple Channel Selector:} We implement a simple delay-based selector, which compares the average packet delay in the last GoP slot for all the $3$ channels and selects the channel with the smallest delay.

\noindent
{\bf (iv)} \emph{Oracle Channel Selector:}
As an absolute upper bound, we consider an oracle channel selector that has a priori knowledge of the PSNR in all available channels. 

The average PSNR achieved by the different selectors in scenarios where nodes are static or move according to a random walk with speed up to 5 m/s is shown in Table~\ref{table:avg_psnr}. It can be seen how the proposed data-driven approach ($51.85$~dB) performs closely to the Oracle channel selector ($54.31$~dB), with a loss of less than $3$~dB, and an improvement of $14$~dB with respect to fixed channel transmission. ABR transmission and Delay-based selection, at $45.27$ and $47.90$~dB, achieve intermediate performance. When considering high mobility, we note a reduction in the achievable average PSNR. However, also in this case the proposed predictive channel selector outperforms other options and performs close to the oracle selector.

\begin{comment}

\begin{figure}[t]
\begin{center}
        \includegraphics[width=0.7\columnwidth]{images/single_ch_vs_dash_vs_ch_selection_4.pdf}
        \caption{Temporal pattern of PSNR achieved by the different transmission strategies.}
        \label{fig:online_temporal}
\end{center}
\end{figure}

\end{comment}

\begin{table}[!t]
\centering
\vspace{2mm}
\begin{tabularx}{\linewidth}{|L|L|L|L|L|L|} 
\hline
%\phantom{ccccc}
\textbf{PSNR (dB)} & \textbf{Fixed Channel} (lower bound) &  \textbf{Oracle Channel Selector} (Upper bound) & \textbf{ABR with full knowledge} & \textbf{Delay-based Simple selector} &  \textbf{Probe-based predictive selector}\\ 
    \hline
\centering No Mobility & \phantom{ccccc}~37.52 &
\phantom{ccccc}~54.31 &
\phantom{ccccc}~45.27 &  \phantom{ccccc}~47.90 & \cellcolor{blue!25}
\phantom{ccccc}~~51.85\\
\hline
\centering With Mobility & \phantom{ccccc}~35.24 &
\phantom{ccccc}~49.13&
\phantom{ccccc}~42.54 & 
 \phantom{ccccc}~45.31 &
 \cellcolor{blue!25}
 \phantom{ccccc}~~48.33\\
    \hline
\end{tabularx}
\vspace{2mm}
\caption{Average PSNR comparisons in different real-time video streaming approaches.}
\vspace{-4mm}
\label{table:avg_psnr}
\end{table}
%\vspace{-4mm}

\begin{figure}[t]
\begin{center}
        \includegraphics[width=0.9\columnwidth]{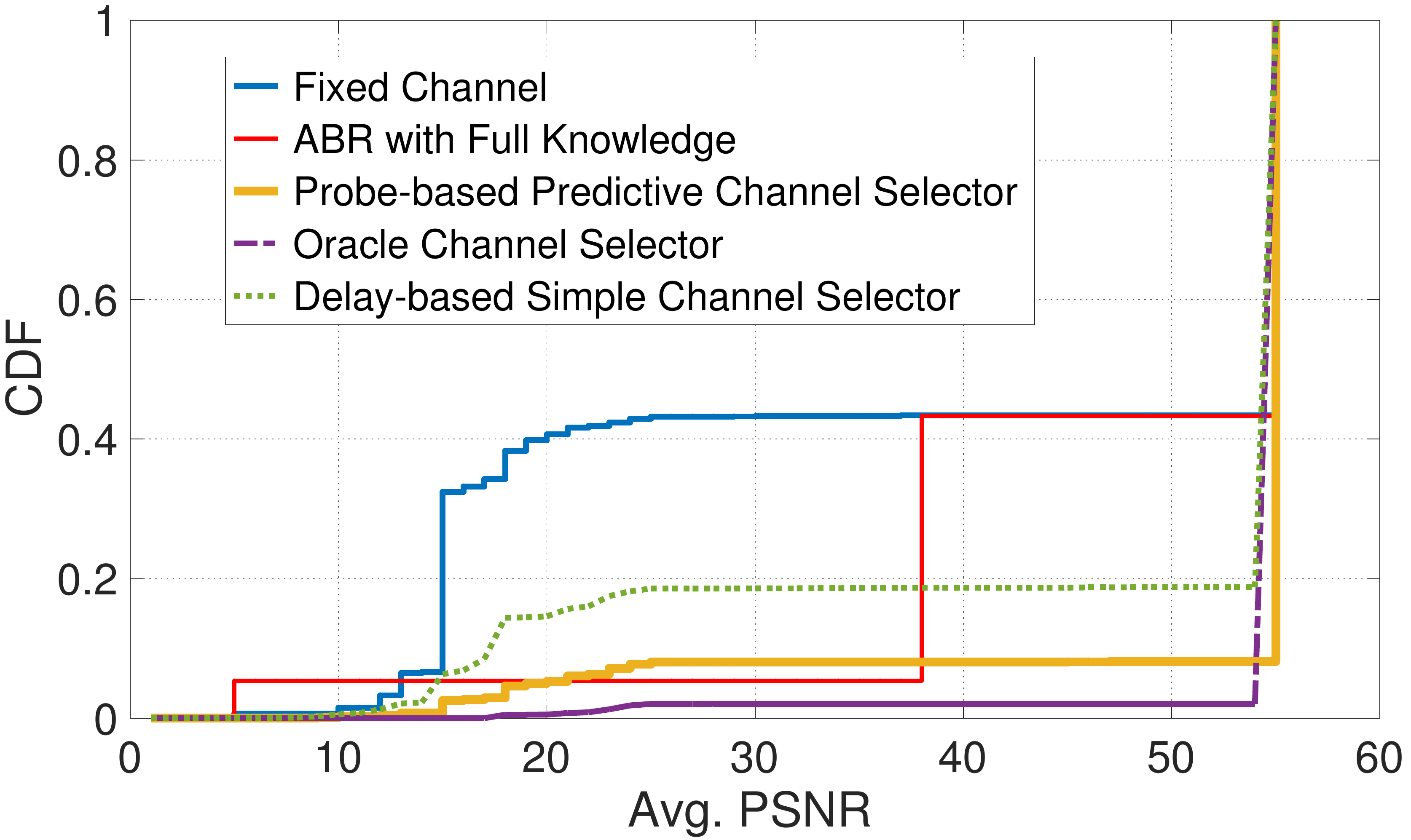}
        \caption{CDF of PSNR for different transmission strategies.}
        \label{fig:psnr_cdf}
        \vspace{-8mm}
\end{center}
\end{figure}

The CDF of the PSNR is shown in
Figure~\ref{fig:psnr_cdf}. Transmission over a fixed channel results into a widely spread distribution of PSNR, which may impair the ability of edge server to perform analysis. The probe-based classifier is capable of eliminating most low and mid-range PSNRs, although prediction errors result in a larger probability of mid-range PSNR than that achieved by the oracle. ABR transmission efficiently removes most low-range PSNRs, but the CDF has a sharp increase at $40$~dBs, emphasizing the limits of adaptation against bad channel conditions, where the compression necessary to effectively deliver the packets produces a perceivable degradation.

%\input{eval.tex}

%\input{related.tex}

%%%%%%%%%%%%%%%%%%%%%%%%%%%%%%%%%%%%%%%%%%%%%%%%%%%%%%
%% Conclusions
%%%%%%%%%%%%%%%%%%%%%%%%%%%%%%%%%%%%%%%%%%%%%%%%%%%%%%

\section{Conclusions}
\label{sec:concl}
This paper presented a data-driven form of prediction and channel selection for wireless channels subject to interference and congestion. The core idea is to build classifiers capable of transforming features collected in real-time by wireless nodes into predicted quality of the data transmitted to edge servers. We specifically consider a scenario where a real-time video stream is acquired and transmitted over highly dynamical channels. An extensive discussion on the predictive power of features extracted from the data and probe stream was provided. We measured the performance of classification and channel selection by means of extensive simulations. Results show prediction accuracy of about $90$\% and average PSNR close to that of an oracle predictor with a limited disruption of other data streams' performance.
%\input{conclusions.tex}

%%%%%%%%%%%%%%%%%%%%%%%%%%%%%%%%%%%%%
%%%%%%    THE  BIBLIOGRAPHY    %%%%%%
%%%%%%%%%%%%%%%%%%%%%%%%%%%%%%%%%%%%%

%\bibliographystyle{plain}
\bibliographystyle{IEEEtran}
\bibliography{iot}

\end{document}